\documentclass[letterpaper, 10 pt, conference]{ieeeconf}  

\IEEEoverridecommandlockouts                              

\usepackage{caption}
\usepackage{subcaption}
\usepackage{amsmath,amssymb}
\usepackage{graphicx,epstopdf}
\usepackage{color}
\usepackage{amsfonts}
\usepackage{stfloats}
\usepackage{bm}
\usepackage[symbol]{footmisc}
\usepackage{algorithmic}

\newcommand{\la}{\lambda}
\newcommand*\diff{\mathop{}\!\mathrm{d}}
\newcommand{\R}{\mathbb{R}}

\newcommand{\C}{\mathcal{C}}
\newcommand{\fot}{\frac{1}{2}}

\newcommand{\Q}{\mathcal{Q}}
\newcommand{\mr}{\mathcal{R}}
\newcommand{\pa}{\partial}

\newcommand{\ms}{\mathcal{S}}

\newcommand{\phb}{\bigg(\frac{\partial h}{\partial x_1}\bigg)}
\newcommand{\ph}{\frac{\partial h}{\partial x_1}}

\allowdisplaybreaks[4]

\newif\ifdraft
\drafttrue

\newtheorem{example}{\bfseries Example}
\newtheorem{assumption}{\it Assumption}
\newtheorem{theorem}{\bfseries Theorem}

\newtheorem{remark}{\bfseries Remark}

\newtheorem{problem}{\bfseries Problem}

\makeatletter
\renewcommand\normalsize{%
\@setfontsize\normalsize\@xpt\@xiipt
\abovedisplayskip 3\p@ \@plus2\p@ \@minus1\p@
\abovedisplayshortskip \z@ \@plus2\p@
\belowdisplayshortskip 3\p@ \@plus2\p@ \@minus1\p@
\belowdisplayskip \abovedisplayskip
\let\@listi\@listI}
\makeatother

\title{\LARGE \bf
Disturbance Observer-based Robust Control Barrier Functions 
}

\author{}

\author{Yujie Wang and Xiangru Xu\thanks{Y. Wang and X. Xu are with the Department of Mechanical Engineering, University of Wisconsin-Madison,
        Madison, WI 53706, USA. Email: 
\{yujie.wang, xiangru.xu\}@wisc.edu.}}

\begin{document}
\maketitle

\begin{abstract}
This work presents a safe control design approach that integrates the disturbance observer (DOB) and the control barrier function (CBF) for systems with external disturbances. Different from existing robust CBF results that consider the “worst case” of disturbances, this work utilizes a DOB to estimate and compensate for the disturbances. DOB-CBF-based controllers are constructed with provably safe guarantees by solving convex quadratic programs online, to achieve a better tradeoff between safety and performance. Two types of systems are considered individually depending on the magnitude of the input and disturbance relative degrees. The effectiveness of the proposed methods is illustrated via numerical simulations.

\end{abstract}

\section{Introduction}
\label{sec:introduction}

Control barrier functions (CBFs) have emerged as one powerful tool for ensuring control system safety in the form of set invariance and have been successfully applied to various autonomous and robotic systems \cite{ames2016control}. Nevertheless,  most existing works on CBF-based control design rely on accurate model information and state measurement, which are usually difficult to obtain in practice. To address this problem, various robust CBF approaches were proposed for systems with model/measurement uncertainties and/or external  disturbances \cite{jankovic2018robust,nguyen2021robust,zhang2022control,wang2022observer}. Most of the robust CBF-based methods consider the ``worst case'' of disturbances and design safe controllers that are often unnecessarily conservative. 

Recently, some robust CBF control schemes based on disturbance estimation and compensation techniques were proposed with the goal of reducing the conservatism of the related safe controllers \cite{dacs2022robust,alan2022robust,cheng2019end,zhao2020adaptive,gu2021safety}. For example, in \cite{dacs2022robust,alan2022robust}, a high-gain input disturbance observer was integrated into the CBF framework; in \cite{cheng2019end}, Gaussian processes were employed to estimate the disturbances/uncertainties from data and an end-to-end safe reinforcement learning scheme was developed based on CBFs; in \cite{zhao2020adaptive}, a  piecewise-constant disturbance estimation law was proposed and integrated into the robust CBF framework; in \cite{gu2021safety}, a CBF-based safe control law was designed for autonomous surface vehicle systems based on the fixed-time extended state observer.

Disturbance observer (DOB) is a special class of unknown input observers. DOBs estimate the internal and external disturbances by using identified dynamics and measurable states of plants, and have been widely employed in applications such as robotics, automotive, and power electronics \cite{mohammadi2017nonlinear,sariyildiz2019disturbance,chen2015disturbance}. In contrast to other worst-cased-based robust control schemes, the DOB-based methods aim to attenuate the influence of disturbances by compensating for the disturbances and achieve a better tradeoff between robustness and performance.  The majority of existing DOB-based control schemes focus on systems whose disturbance relative degree is 
higher than or equal to the input relative degree \cite{yang2013static}; however, systems with a lower disturbance relative degree 
are ubiquitous (such as the missile system \cite{yang2013static}, the flexible joint manipulator \cite{ginoya2013sliding}, and the 
PWM-based DC–DC buck power converter 
system \cite{wang2015extended}), 
and various results were recently proposed to design DOBs for such systems.

\begin{figure}[!t]
\centering
\begin{subfigure}{0.5\textwidth}
  \centering
  \includegraphics[width=0.85\textwidth]{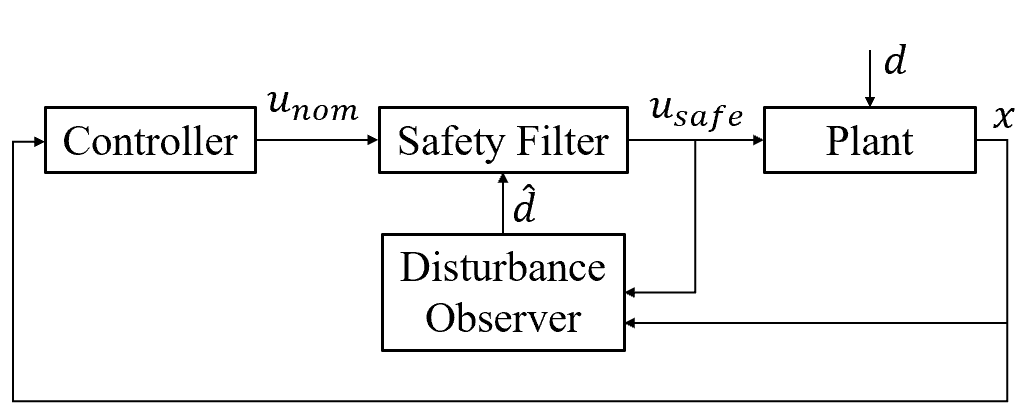}
  \caption{Architecture of the proposed DOB-CBF-QP framework.}
\end{subfigure}
\newline
\begin{subfigure}{0.5\textwidth}
  \centering
  \includegraphics[width=0.85\textwidth]{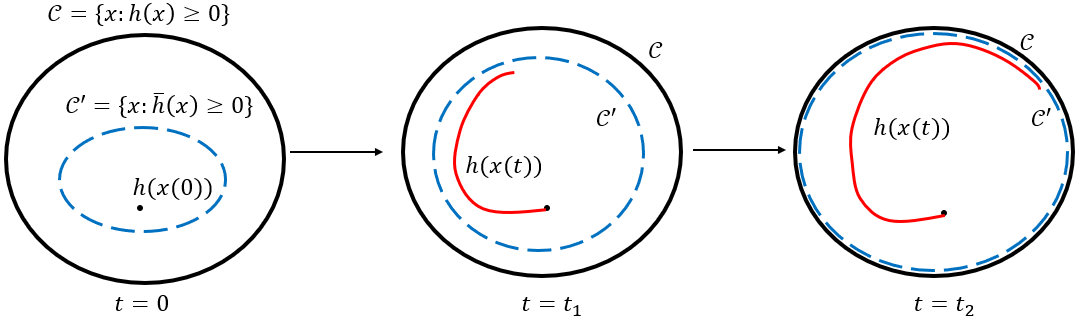}
  \caption{Illustration of the evolution of the safe set $\C'$. }
\end{subfigure}
\caption{As shown in (b), the trajectories of the closed-loop system are guaranteed to stay in a set $\C'$.  
Since the DOB will ensure $\hat d$ converge to $d$, the set $\C'$ 
keeps expanding and can be made arbitrarily close to the original safe set, $\C$, by choosing parameters appropriately. In contrast, robust CBF design based on the ``worst-case'' of disturbances will result in a safe set whose size will shrink as the magnitude of the disturbances become larger.}
    \label{fig0}
\end{figure}

This paper develops a safe control design method that integrates the DOB and the CBF for systems with external
disturbances. As shown in Fig. \ref{fig0}(a), a DOB is introduced to generate an estimate of the disturbance, which is used by the CBF-based safety filter to compensate for the disturbance. By solving a convex quadratic program (QP) online, a safe control law is obtained that can achieve a better tradeoff between safety and performance.  
Specifically, this paper first presents a DOB-CBF-QP-based safe control design method for systems whose input relative degree is not higher than the disturbance relative degree. Compared with existing results, the proposed approach relies on milder assumptions and can provide a robust safety guarantee even if the bound of the disturbances is not exactly known. This paper also establishes two DOB-CBF-QP-based approaches for a class of systems whose input relative degree is lower than the disturbance relative degree, by using the recursive CBF design and the extended DOB techniques, respectively. To the best of our knowledge, this is the first safe control design result for such class of systems.

The remainder of this paper is organized as follows. In Section \ref{sec:preliminary}, preliminaries about CBFs and DOBs are introduced, and the problem  formulation is provided; in Section \ref{sec:main}, the main results of this paper are presented; in Section \ref{sec:simulation}, numerical simulation results are provided to validate the proposed methods; and finally, the conclusion is drawn in Section \ref{sec:conclusion}.

\section{Preliminaries \& Problem Statement}\label{sec:preliminary}
\subsection{Control Barrier Function}\label{sec:relativedegree}
Consider a system 
\begin{equation}
    \dot x = f(x)+g_1(x) u+g_2(x) d(t),\label{controlaffine}
\end{equation}
where $x\in\R^n$ is the state, $u\in\R^m$ 
is the control input, $f: \R^n\to\R^n$, $g_1: \R^n\to\R^{n\times m}$ and $g_2: \R^n\to\R^{n\times q}$ are known and sufficiently smooth functions, and $d(t):\R\to \R^q$ represents the unknown external disturbance. 

Suppose that $h(x):\R^n\to\R$ is a sufficiently smooth function associated with system \eqref{controlaffine}. If $m=q=1$, then system \eqref{controlaffine} is said to have an \emph{input relative degree} of $r (1\leq r\leq n)$ in a region $\mathcal{D}$ if 
\begin{align*}
&L_{g_1}L_f^{r-1}h(x_0)\neq 0,\;L_{g_1}L_f^ih(x)=0, i=1,\cdots,r-2,
\end{align*}
for all $x\in\mathcal{D}$, and an \emph{disturbance relative degree} of $\rho (1\leq \rho\leq n)$ if 
\begin{align*}
&L_{g_2}L_f^{\rho-1}h(x_0)\neq 0,\; L_{g_2}L_f^jh(x)=0,j=1,\cdots,\rho-2,
\end{align*}
for all $x\in\mathcal{D}$, where $L_{g_s}L_f^th$ are Lie derivatives \cite{isidori1985nonlinear,yang2013static,khalil2002nonlinear}. 
If $m>1$ and/or $q>1$, then the vector input and disturbance relative degrees can be similarly defined following \cite[Section 5]{isidori1985nonlinear}.

In \cite{Xu2015ADHS} and \cite{xu2018constrained}, (zeroing) CBFs with different relative degrees were introduced for disturbance-free systems. Specifically, 
given system \eqref{controlaffine} with $g_2=0$ and a safe set $\mathcal{C} \subset \R^n$ defined by 
\begin{equation}\label{setc}
    \mathcal{C} = \{ x \in \R^n : h(x) \geq 0\},
\end{equation}
the function $h$ is called a CBF of (input) relative degree 1 
if 
\begin{align}\label{ineqZCBF}
& \sup_{u }  \left[ L_f h(x) + L_{g_1} h(x) u + \gamma h(x)\right] \geq 0
\end{align}
for all $x\in\R^n$, where $\gamma>0$ is a given positive constant. It was proven in  \cite{Xu2015ADHS} that if 
$h(x(0))> 0$, then any Lipschitz continuous control input $u(x)\in \{ u \mid L_f h(x) + L_{g_1} h(x) u + \gamma h(x) \geq 0\}$ will ensure the forward invariance of $\C$. 
Similarly, the function $h$ is called a CBF of (input) relative degree $r(r\geq 2)$ if there exists ${\bf a}\in\R^r$, such that  
\begin{align}\label{ineq:ZCBF2}
	& \sup_{u }  [L_{g_1} L_f^{r-1}h(x)u \!+\!L_f^rh(x)\!+\!{\bf a}^\top\eta(x) ] \geq 0, 
\end{align}
for all $x\in\R^n$, where $\eta(x)=[L_f^{r-1}h, L_f^{r-2}h,...,h]^{\top}\in\R^r$, and ${\bf a}=[a_1,...,a_r]^{\top}\in\R^r$ is a set of parameters chosen such that the roots of $\la^r+a_1\la^{r-1}+...+a_{r-1}\la+a_r=0$ are all negative reals $-\la_1,...,-\la_r<0$.  The functions $s_k(x(t))$ for $k=0,1,...,r$ are defined recursively as
\begin{align}\label{liftedh}
s_0(x(t))&=h(x(t)),\;s_{k}(x(t))=\bigg(\frac{\diff}{\diff t}+\la_k\bigg)\circ s_{k-1}.
\end{align}
If 
$s_k(x(0))> 0$ for $k=0,1,...,r-1$, then any Lipschitz continuous control input $u(x)\in \{ u \mid L_{g_1} L_f^{r-1}h(x)u +L_f^rh(x)+{\bf a}^\top\eta(x) \geq 0\}$ will ensure the forward invariance of $\C$ \cite{nguyen2016exponential,xu2018constrained}. 

If the system \eqref{controlaffine} is subject to bounded disturbance, i.e., $g_2\neq 0$ and $\|d(t)\|\leq w, \forall t>0$, where $w>0$ denotes the disturbance bound, then the function $h$ is called a robust CBF of relative degree 1 if 
\begin{align}\label{robustZCBF}
& \sup_{u}  \left[ L_f h(x)\! +\! L_{g_1} h(x) u \!-\!\|L_{g_2}h\|w \!+\! \gamma h(x)\right]\! \geq\! 0
\end{align}
for all $x\in\R^n$ and for all $\|d\|\leq w$, where $\gamma>0$ is a given positive constant \cite{jankovic2018robust,nguyen2021robust}. The robust CBF with a higher relative degree can be defined similarly as above.

\subsection{Disturbance Observer \& Extended Disturbance Observer}\label{subsection:dob}

\subsubsection{Disturbance Observer}
We follow the results of \cite{chen2004disturbance} to introduce the DOB that will be used for safe control design in Section \ref{sec:main}. A standard assumption for DOB is given first.
\begin{assumption}\label{assumption1}
The disturbance $d(t)$ and its derivative $\dot d(t)$ are bounded by known positive constants, i.e., $\|d(t)\|\leq \omega_0$ and $\|\dot{d}(t)\|\leq \omega_1, \forall t>0$ where $\omega_0>0$ and $\omega_1>0$.
\end{assumption}

Given system \eqref{controlaffine}, we consider the following DOB:
\begin{equation}
    \begin{cases}
    \hat d = z + \alpha p,\\
     \dot z = -\alpha L_d
    (f+g_1u+g_2\hat d),\\
    \end{cases}\label{adaptivelaw}
\end{equation}
where $\hat d$ is the disturbance estimation,
$L_d(x)$ is the observer gain satisfying 
$
-x^\top L_d g_2 x\leq - x^\top x
$
for any $x$ (e.g., 
$L_d=- (g_2^\top g_2)^{-1}g_2^\top$ if $g_2(x)$ has a full column rank), $\alpha>0$ is a positive \emph{tuning parameter}, and $p(x)$ is a function satisfying $\frac{\pa p}{\pa x}=L_d(x)$. The design of $p(x)$ and $L_d$ is non-trivial and problem-specific; see \cite{li2014disturbance,mohammadi2017nonlinear,chen2015disturbance,sariyildiz2019disturbance} for more details.

Define the \emph{disturbance estimation error} as
\begin{align}\label{estimateerror}
e_d = \hat d-d.
\end{align} 
Then,  $\dot e_d = \dot{\hat{d}} -\dot d =\dot{z}+\alpha\frac{\pa p}{\pa x}\dot x -\dot d = \dot{z}+\alpha L_d\dot x -\dot d.$
Substituting \eqref{controlaffine} and \eqref{adaptivelaw} into this equality yields
\begin{equation}
    \dot e_d=-\alpha L_d g_2 e_d - \dot d.\label{ed3}
\end{equation}
Choose a Lyapunov candidate function 
$V_1 = \fot\| e_d\|^2$.
Invoking \eqref{ed3}, Assumption \ref{assumption1}, and the definition of $L_d$, we have
\begin{align}
\dot V_1 &=- e_d^\top \alpha L_d g_2 e_d -e_d^\top \dot{d} \leq -\alpha  e_d^\top e_d+ \omega_1 \|e_d\| \nonumber\\
&\leq-2\bigg(\alpha -\frac{\nu_1}{2}\bigg)V_1+\frac{\omega_1^2}{2\nu_1}=-2\kappa V_1+\frac{\omega_1^2}{2\nu_1}\label{dotv}
\end{align}
where $\kappa\triangleq\alpha-\frac{\nu_1}{2}$, $\nu_1$ is a constant satisfying $0<\nu_1<2\alpha$, and the second inequality is from the fact that $\omega_1\|e_d\|\leq \frac{\nu_1}{2} \|e_d\|^2+\frac{1}{2\nu_1} \omega_1^2$. 
Recalling comparison lemma \cite{khalil2002nonlinear}, we have
\begin{equation}
    \|e_d(t)\| \leq \sqrt{\frac{2\nu_1\kappa \|e_d(0)\|^2 e^{-2\kappa t} + \omega_1^2(1-e^{-2\kappa t})}{2\nu_1\kappa}}.\label{errorbound}
\end{equation}
Form \eqref{errorbound} one can see the disturbance estimation error $e_d$ is uniformly ultimately bounded. 



\subsubsection{Extended Disturbance Observer}\label{sec:edob}

As a generalization of DOB, the extended DOB was proposed in \cite{ginoya2013sliding} to estimate the high order derivatives of disturbances.   
Consider the following system 
\begin{equation}
    \dot x = f(x,u) + d(t)
\end{equation}
where $x\in\R^n$ is the state, $u\in\R^m$ is the control input, and $d: \R\to\R^n$ is the external disturbance. Similar to Assumption \ref{assumption1}, we assume $d$ and its derivatives are bounded.

\begin{assumption}\label{assumption2}
The disturbance $d$ and its derivatives $d^{(1)},...,d^{(r)}$, where $r$ is a fixed positive integer, are bounded by some known constants, i.e., $\|d_E(t)\|\leq \xi_0$ and $\|d^{(r+1)}(t)\|\leq \xi_1$ for any $t>0$, where $\xi_0>0,\xi_1>0$, and
\begin{align}\label{deq}
d_E(t)\triangleq\begin{bmatrix}
d^{(0)\top} & d^{(1) \top}&\cdots&d^{(r)\top}
\end{bmatrix}^\top    
\end{align}
with $d^{(0)}=d$.
\end{assumption}

Consider the following extended DOB as in \cite{ginoya2013sliding}:
\begin{IEEEeqnarray}{rCl}
\IEEEyesnumber\label{edob:both} 
\IEEEyessubnumber
\hat d^{(i)} &=& p_i+l_ix,\label{edob:sub1}\\
\IEEEyessubnumber \dot p_i &=& - l_i(f+\hat d^{(0)})+\hat d^{(i+1)}, \label{edob:sub2}
\end{IEEEeqnarray}
where $\hat d^{(i)}$ denotes the estimate of $d^{(i)}$, $l_i$ is a tuning parameter,  $i=0,1,\cdots,p$, and $\hat d^{(r+1)}=0$. Define $\hat d_E(t)=\begin{bmatrix}
\hat d^{(0)\top} &\hat d^{(1) \top}&\cdots&\hat d^{(r)\top}
\end{bmatrix}^\top$ and the estimation error
\begin{equation}
    \tilde e_d = \hat d_E-d_E.\label{errored}
\end{equation}
Equations \eqref{edob:both} can be written compactly as
\begin{equation}
    \dot {\tilde e}_{d} = A{\tilde e}_{d} + Bd^{(r+1)},
\end{equation}
where
\begin{equation*}
    A = \begin{bmatrix}
    -l_0&1&0&\cdots&0\\
    -l_1&0&1&\cdots&0\\
    \vdots&\vdots&\vdots&\vdots&\vdots\\
    -l_{r-1}&0&0&\cdots&1\\
    -l_{r}&0&0&\cdots&0\\
    \end{bmatrix}, \ 
    B = \begin{bmatrix}
    0\\0\\\vdots\\0\\-1
    \end{bmatrix},
\end{equation*}
and $l_0, \cdots, l_p$ are selected such that $\lambda_{\min} (A)<0$. Define a Lyapunov candidate function as $V_2=\fot \|\tilde e_d\|^2$. Then,
\begin{align}
    \dot V_2 &= \tilde e_d^\top A \tilde e_d \!+\! \tilde e_d^\top B d^{(r+1)}\!\leq\! \lambda_{\min} (A)\|\tilde e_d\|^2\!+\!\xi_1 \|\tilde e_d\|\nonumber\\
    &\leq\!\left(\lambda_{\min} (A)\!+\!\frac{\nu_2}{2}\right) \|\tilde e_d\|^2\!+\!\frac{\xi_1^2}{2\nu_2}\!=\!-2\kappa_EV_2\!+\!\frac{\xi_1^2}{2\nu_2},\label{edobdotv}
\end{align}
where $\kappa_E\triangleq-\lambda_{\min} (A)-\frac{\nu_2}{2}$ and $\nu_2$ is a constant satisfying $0<\nu_2<-2\lambda_{\min} (A)$. Similar to \eqref{errorbound}, 
\begin{equation}
    \|\tilde e_d(t)\|\!\leq\! \sqrt{\frac{2\nu_2\kappa_E \|\tilde e_d(0)\|^2e^{-2\kappa_E t}\!+\!\xi_1^2(1\!-\!e^{-2\kappa_E t})}{2\nu_2\kappa_E}}.\label{bounded}
\end{equation}
It can be seen the $\tilde e_d$ is uniformly ultimately bounded. 

\subsection{Problem Statement}\label{sec:problemformulation}
In this paper, we will consider the DOB-CBF-based safe control design problem for two types of systems individually depending on the magnitude of the input and disturbance relative degrees. In the first problem, the system  has an input relative degree not higher than its disturbance relative degree.


\begin{problem}\label{prob1}
Given system \eqref{controlaffine} whose input relative degree is not higher than its disturbance relative degree, the safe set $\mathcal{C}$ defined in \eqref{setc}, and the DOB given in \eqref{adaptivelaw}, design a controller $u(x,\hat d)$ such that the closed-loop system is safe with respect to $\mathcal{C}$, i.e., $h(x(t))\geq 0$ for all $t\geq 0$ provided that $x(0)\in\C$.
\end{problem}

For the second problem, we consider the following system with a mismatched disturbance:
\begin{equation}\label{system1dmismatched}
\begin{split}
    \dot x_1 &= f_1(\bar x_2) + d,\\
    \dot x_2 &= f_2(\bar x_3),\\
    &\cdots\\
    \dot x_n &= f_n(\bar x_n)+g(\bar x_n)u,
\end{split}
\end{equation}
where $x_i \in\R$ is the state, $u\in\R$ is the control input, $d\in\R$ is the mismatched disturbance, and $\bar x_i= [x_1 \ x_2\ \cdots \ x_i]^\top\in\R^{i}$, $i=1,2,\cdots,n$.  The safe set for such a system is given as
\begin{equation}
    \tilde{\C} = \{(x_1,\cdots,x_n)\in\R^n: h(x_1)\geq 0 \},\label{setcmismatched}
\end{equation}
where $h:\R\to\R$ is a $C^n$ function.
Clearly, for system \eqref{system1dmismatched} with the output function $h$ defined in \eqref{setcmismatched}, its disturbance relative degree  is lower than its input relative degree. 
\begin{problem}\label{prob2}
Given system \eqref{system1dmismatched} and the safe set ${\mathcal{\tilde C}}$ defined in \eqref{setcmismatched}, design a DOB with the estimated disturbance $\hat{d}$ and a controller $u(x,\hat d)$ such that the closed-loop system is safe with respect to ${\mathcal{\tilde C}}$ , i.e., $h(x_1(t))\geq 0$ for all $t\geq 0$ provided that $\bar x_n(0)\in{\mathcal{\tilde C}}$.
\end{problem}


\begin{remark}\label{remark:matched}
In this paper we only consider the single-input-single-output system \eqref{system1dmismatched} with one disturbance due to the page limit. However, the proposed methods can be readily extended to more general systems, such as the nonlinear missile model studied in Example \ref{example2}. The detailed design procedure will be given in our future work.

\end{remark}

\section{Main Results}
\label{sec:main}

In this section, the main results of this paper are presented. In Section \ref{sec:dobmatch}, a DOB-CBF-based  QP is proposed for the system whose input relative degree is not higher than its disturbance relative degree. In Section \ref{sec:mismatcheddobre}, two
DOB-CBF-based QPs are developed for the system whose input relative degree is higher than its disturbance relative degree, by using recursive CBF design and extended DOB techniques, respectively; compared with the first approach, the second one tends to have less conservative safe controller in simulation but it requires more restrictive assumptions (see simulation examples in Section \ref{sec:simulation}).


\subsection{DOB-CBF-QP for Solving Problem 1}
\label{sec:dobmatch}
 In this subsection, we will present the DOB-CBF-based safe control design method to solve Problem \ref{prob1}. We will first consider the simple case where the CBF $h$ has an input relative degree 1, and then generalize the result to the case where $h$ has a higher input relative degree $r(r>1)$.

The following result is the first main result of this work for the CBF $h$ with an input relative degree 1.
\begin{theorem}\label{theorem2}
Consider the system \eqref{controlaffine}, the safe set $\mathcal{C}$ defined in \eqref{setc}, and the DOB given in \eqref{adaptivelaw} with $\hat d(0)=0$. Suppose that Assumption \ref{assumption1} holds, $h$ has an input relative degree 1, $h(x(0))> 0$, and there exist positive constants $\gamma>0$, $\alpha>\frac{\gamma+\nu_1}{2}$, $\beta > \frac{\|e_d(0)\|^2}{2h(x(0))}$, such that
\begin{IEEEeqnarray}{rCl}
\sup_u &\bigg[L_f h- \|L_{g_2} h\|\chi -\frac{\omega_1^2}{2\nu_1\beta}-\frac{\beta \|L_{g_2}h\|^2}{4\alpha-2\nu_1-2\gamma} \nonumber\\
     &+\gamma  h+L_{g_1}hu \bigg]\geq 0, \label{cbfcondition}
\end{IEEEeqnarray}
where $\chi=\omega_0+\sqrt{\omega_0^2+\frac{\omega_1^2}{2\nu_1\kappa}}$. Then, any Lipschitz continuous controller  
$$
u(x) \in K_{DOB}(x,\hat{d})\triangleq\left\{u\mid \psi_{0}+\psi_{1} u\geq 0\right\}
$$  
where
\begin{subequations}\label{eqpsi}
\begin{align}
\psi_{0}&=L_f h\!+\! L_{g_2} h \hat d\!-\!\frac{\omega_1^2}{2\nu_1\beta}\!-\!\frac{\beta \|L_{g_2}h\|^2}{4\alpha\!-\!2\nu_1\!-\!2\gamma}\!+\!\gamma  h,\label{eqpsia}\\
\psi_{1}&=L_{g_1} h,    
\end{align}
\end{subequations}
will guarantee $h(x(t))\geq 0$ for all $t\geq 0$.
\end{theorem}
\begin{proof} If $\hat d(0) = 0$, then $\hat d(t)$ satisfies
$\|\hat d(t)\|=\|d(t)+e_d(t)\|\leq \chi$ from \eqref{errorbound}. 
Define a new candidate CBF $\bar h$ as
\begin{equation}
    \bar{h}(x(t),t) = \beta h -\fot e_d^\top e_d\label{barh1}
\end{equation}
where $e_d$ is defined in \eqref{estimateerror}. 
From \eqref{barh1}, it can be seen $\bar h\geq 0$ implies $h\geq \frac{\|e_d\|^2}{2\beta}\geq 0$. 
Since $\beta>\frac{\|e_d(0)\|^2}{2h(x(0))}$, one can verify
$\bar{h}(x(0),0) >  0$.
Moreover, $\dot{\bar{h}}$ satisfies
\begin{IEEEeqnarray}{rCl}
  \hspace{-3mm} \dot{\bar{h}} 
   &\stackrel{\eqref{dotv}}{\geq}& \beta( L_fh+L_{g_1} h u+L_{g_2}h d )+\kappa e_d^\top e_d-\frac{\omega_1^2}{2\nu_1}\nonumber\\
   &=& \beta (L_fh+ L_{g_1} h u+L_{g_2}h \hat d -L_{g_2}h e_d)+\frac{\gamma}{2}e_d^\top e_d \nonumber\\
   &&+\bigg(\alpha-\frac{\nu_1}{2}-\frac{\gamma}{2}\bigg)e_d^\top e_d-\frac{\omega_1^2}{2\nu_1}  \nonumber\\
   &=& \beta (L_fh+ L_{g_1} h u+ L_{g_2}h \hat d) -\frac{\omega_1^2}{2\nu_1}-\frac{\beta^2 \|L_{g_2}h\|^2 }{4\alpha-2\nu_1-2\gamma}
   \nonumber\\
   && + \bigg\|\sqrt{\alpha-\frac{\nu_1}{2}-\frac{\gamma}{2}}e_d^\top- \frac{\beta L_{g_2}h}{2\sqrt{\alpha-\frac{\nu_1}{2}-\frac{\gamma}{2}}}\bigg\|^2_2+\frac{\gamma}{2}e_d^\top e_d \nonumber\\
   &\geq &\beta \left(\psi_{0}+\psi_{1} u-\gamma h\right) +\frac{\gamma}{2}e_d^\top e_d.\nonumber
\end{IEEEeqnarray}
Therefore, any $u\in K_{DOB}$  yields
\begin{equation*}
    \dot{\bar{h}} \geq -\gamma \beta h +\frac{\gamma}{2}e_d^\top e_d  \geq -\gamma \bigg(\beta h - \fot e_d^\top e_d\bigg) = -\gamma \bar h,
\end{equation*}
which implies that $\bar h(x(t),t)\geq 0$ for all $t\geq 0$ because $\bar h(x(0),0)\geq 0$. Therefore, $h(x(t))\geq  0, \forall t>0$.
\end{proof}

The safe controller proposed in Theorem \ref{theorem2} is obtained by solving the following DOB-CBF-QP:%
\begin{align}
\min_{u} \quad & \|u-u_{nom}\|^2\label{cbfQP1}\\
\textrm{s.t.} \quad & \psi_0+\psi_1 u\geq 0, \nonumber\\
&\mbox{DOB given in} \; \eqref{adaptivelaw},\nonumber
\end{align}
where $\psi_0,\psi_1$ are given in \eqref{eqpsi} and $u_{nom}$ is any given nominal control law.
\begin{remark}
The proof of Theorem  \ref{theorem2} reveals that, by ensuring $\bar h(t)\geq 0$, $x(t)$ is restricted to stay in a set defined by $\C'(t)\triangleq\{x\mid h(x)\geq \frac{\|e_d(t)\|^2}{2\beta}\}$. According to \eqref{errorbound}, the ultimate bound of $\|e_d(t)\|$ is $\frac{\omega_1}{\sqrt{\nu_1(2\alpha-\nu_1)}}$; therefore, $\C'(t)$ will eventually converge to the set 
$\C'(\infty)\triangleq\{x\mid h(x)\geq \frac{\omega_1^2}{2\beta\nu_1(2\alpha-\nu_1)}\}$. 
By choosing the parameters $\alpha,\nu_1,\beta$ appropriately (e.g., choose $\alpha$ or $\beta$ large enough with other parameters fixed), the set $\C'(\infty)$ can be made arbitrarily close to the original safe set $\C$ despite the unknown disturbance; that is, the system trajectory is allowed to  approach arbitrarily close to the boundary of $\mathcal{C}$ (see also Fig. \ref{fig0} (b)).

\end{remark}

\begin{remark}\label{remark6}
If $K_{DOB}(x,\hat{d})$ is modified by dropping the term $\frac{\omega_1^2}{2\nu_1\beta}$ in \eqref{eqpsia}, then any $u(x) \in K_{DOB}(x,\hat{d})$ yields $\dot{\bar{h}} \geq -\gamma \bar h -\frac{\omega_1^2}{2\nu_1}$, which implies input-to-state safety of the system \cite{kolathaya2018input,lyu2022small}. Therefore, even when $\omega_1$ is not exactly known, the proposed controller may be modified to serve as an input-to-state safe controller; in contrast, the DOB-CBF-based method in \cite{dacs2022robust} requires the exact value of $\omega_1$ in the control design. 
\end{remark}

Now  we consider the case where a CBF $h$ has a higher input relative degree $r(r>1)$. To simplify the expression, we assume $u$ and $d$ in \eqref{controlaffine} are scalars; however, the proposed method can be readily extended to the case where $u$ and $d$ are vectors. In Section \ref{sec:simulation}, we will present a robot manipulator example whose input and disturbance are both vectors.



\begin{theorem}\label{theorem3}
Consider the system \eqref{controlaffine} with  dimensions $m=q=1$,  the safe set $\C$ defined in \eqref{setc}, and the DOB given in \eqref{adaptivelaw} with $\hat d(0)=0$. Suppose that Assumption \ref{assumption1} holds, $h$ has an input relative degree $r(r>1)$, $s_k(x(0))> 0$ for $k=0,1,...,r-1$, where $s_k(x(t))$ is defined in \eqref{liftedh}, and there exist ${\bf a}\in\R^{r}$,
$\beta> \frac{\|e_d(0)\|^2}{2s_{r-1}(x(0))}$, $\alpha>\frac{\lambda_r+\nu_1}{2}$,
such that
\begin{IEEEeqnarray}{rCl}
   \sup_{u}& \bigg[ L_f^{r} h-\| L_{g_2}L_f^{r-1} h\|\chi -\frac{\omega_1^2}{2\nu_1\beta}\!-\!\frac{\beta \|L_{g_2}L_f^{r-1}h\|^2}{4\alpha\!-\!2\lambda_{r}\!-\!2\nu}\nonumber\\
   &+{\bf a}^\top\eta(x)+L_{g_1}L_f^{r-1} hu\bigg]\geq 0, \label{relativercbfcondition}
\end{IEEEeqnarray}
where ${\bf a}$, $\eta$ are defined in \eqref{ineq:ZCBF2} and $\chi=\omega_0+\sqrt{\omega_0^2+\frac{\omega_1^2}{2\nu_1\kappa}}$.
Then any Lipschitz continuous controller  
$$
u(x) \in K_{DOB}^r(x,\hat{d})\triangleq\left\{u\mid \psi_{0}^r+\psi_{1}^r u\geq 0\right\}
$$  
where
\begin{subequations}\label{kcbfr}
\begin{align}
\psi_{0}^r\!&=\!L_f^{r} h\!+\!  L_{g_2}L_f^{r-1} h \hat d\!-\!\frac{\omega_1^2}{2\nu_1\beta}\!-\!\frac{\beta \|L_{g_2}L_f^{r-1}h\|^2}{4\alpha\!-\!2\lambda_{r}\!-\!2\nu}\! +\!{\bf a}^\top\eta(x),\\
\psi_{1}^r\!&=\!L_{g_1}L_f^{r-1} h ,    
\end{align}
\end{subequations}
will guarantee $h(x(t))\geq 0$ for all $t\geq 0$. 
\end{theorem}

\begin{proof}
We define a  new CBF candidate as
\begin{equation}
    \bar{h}^r(x,t)= \beta s_{r-1}(x) -\fot e_d^\top e_d.
\end{equation}
It can be easily verified that selecting $u\in K_{DOB}^r$ gives
$\dot{\bar{h}}^r \geq -\lambda_r \bar h^r$,
and $\beta> \frac{\|e_d(0)\|^2}{2s_{r-1}(x(0))}$ indicates $\bar h^r(x(0))\geq 0$. Therefore, one can see that $\bar h^r\geq 0$, which indicates $h(x_1(t))\geq 0$ for any $t>0$ because $s_k(x(0))>0$, $k=0,1,\cdots,r-1$ \cite{xu2018constrained}. 
\end{proof}

The safe controller proposed in Theorem \ref{theorem3} is obtained by solving the following DOB-CBF-QP:%
\begin{align}
\min_{u} \quad & \|u-u_{nom}\|^2\label{cbfQP1r}\\
\textrm{s.t.} \quad & \psi_0^r+\psi_1^r u\geq 0, \nonumber\\
&\mbox{DOB given in} \; \eqref{adaptivelaw},\nonumber
\end{align}
where $\psi_0^r,\psi_1^r$ are given in \eqref{kcbfr} and $u_{nom}$ is any given nominal control law.


\subsection{DOB-CBF-QP for Solving Problem \ref{prob2}}\label{sec:mismatcheddobre}


In this subsection, we will present two DOB-CBF-based safe control design methods to solve Problem \ref{prob2}. The first method relies on a DOB and recursive CBF design, while the second method is based on an extended DOB. We consider the system \eqref{system1dmismatched} and the safe set defined in \eqref{setcmismatched}. 

\subsubsection{DOB and Recursive CBF-Based Method}
In this method, we design the following DOB as given in \eqref{adaptivelaw}:
\begin{equation}\label{dobmismatched1d}
\begin{split}
\hat d &= z + \alpha x_1, \\
\dot z &= -\alpha(f_1(x_1,x_2) +\hat d),
\end{split}
\end{equation}
From \eqref{dobmismatched1d} we have $\dot{\hat{d}}=-\alpha e_d$, where $e_d$ is defined in \eqref{estimateerror}. 
Define a set of functions $\bar h_i(\bar x_{i+1},t)$, $i=0,1,\cdots,n-1$ as follows:
\begin{align}
\bar h_0(x_1,t)&\!=\!h(x_1),\\
\bar h_i(\bar x_{i+1},t) &\!=\! \Q_i(\bar x_{i+1},\hat d)-\beta_i V_1,\label{barhi}
\end{align}
where $V_1=\fot e_d^2$, $\Q_i(\bar x_{i+1},\hat d), i=1,2,\cdots,n-1$, is defined recursively as
\begin{IEEEeqnarray}{rCl}
\IEEEyesnumber\label{mathcalq:both}
\IEEEyessubnumber \label{mathcalq:1}
\Q_1(\bar x_2,\hat d)&=& \ph (f_1+\hat d) -\frac{1}{2\beta_1}\phb^2+\lambda_1 h,\\
\Q_i(\bar x_{i+1},\hat d) &=& \frac{\pa \Q_{i-1}}{\pa x_1}(f_1+\hat d)+\sum_{j=2}^{i}\frac{\pa \Q_{i-1}}{\pa x_j}f_j+\lambda_{i}\Q_{i-1}\nonumber\\
\IEEEyessubnumber \label{mathcalq:i}
    &&-\frac{\left(\alpha\frac{\pa \Q_{i-1}}{\pa \hat d}+\frac{\pa \Q_{i-1}}{\pa x_1} \right)^2}{\beta_{i-1}\left(4\alpha\!-\!2\nu_1\!-\!2\lambda_{i}\right)}\!-\!\frac{\beta_{i-1}\omega_1^2}{2\nu_1},
\end{IEEEeqnarray}
and $0<\lambda_i<2\alpha-\nu_1$, $\beta_i>0$ are tuning parameters. Define
\begin{align*}
\mathcal{P}(\bar x_n,
\hat d)&=\frac{\pa \Q_{n-1}}{\pa x_1}(f_1+\hat d)+\sum_{j=2}^{n-1}\frac{\pa \Q_{n-1}}{\pa x_j}f_j-\frac{\beta_{n-1}\omega_1^2}{2\nu_1}\\
&-\frac{\left(\!\alpha\frac{\pa \Q_{n-1}}{\pa \hat d}\!+\!\frac{\pa \Q_{n-1}}{\pa x_1}\!\right)^2}{\beta_{n-1}\left(4\alpha\!-\!2\nu_1\!-\!2\gamma\right)}.    
\end{align*}
It is easy to verify that $\frac{\pa \Q_{n-1}}{\pa x_n}$ is independent of $\hat d$. Meanwhile, if $x_1,\cdots,x_n$ are fixed, $\mathcal{P}(\bar x_n,
\hat d)$ and $\Q_{n-1}(\bar x_n,
\hat d)$ can be represented as polynomials of $\hat d$. Hence, one can see that there exist functions $\mathcal{F}_i(\bar x_n), i=0,1,\cdots,n-1$ and $\mathcal{G}_j(\bar x_n), j=0,1,\cdots,n$ such that $\Q_{n-1}(\bar x_n,\hat d)$ and $\mathcal{P}(\bar x_n,\hat d)$ can be expressed as
\begin{IEEEeqnarray}{rCl}
\IEEEyesnumber \label{dothpoly}
\IEEEyessubnumber
   \Q_{n-1}(\bar x_n,\hat d)&=&\mathcal{F}_0(\bar x_n)+\sum_{i=1}^{n-1} \mathcal{F}_i(\bar x_n)\hat d^i,\\
\IEEEyessubnumber
   \mathcal{P}(\bar x_n,\hat d) &=& \mathcal{G}_0(\bar x_n)+\sum_{i=1}^{n} \mathcal{G}_i(\bar x_n)\hat d^i.
\end{IEEEeqnarray}


\begin{theorem}\label{theorem5}
Consider the system \eqref{system1dmismatched}, the safe set $\tilde\C$ defined in \eqref{setcmismatched}, and the DOB given in \eqref{dobmismatched1d} with $\hat d(0)=0$. Suppose that Assumption \ref{assumption1} holds, $\bar h_i(\bar x_{i+1}(0),0)\geq 0$, $i=0,1,\cdots,n-1$, where $\bar h_i$ is defined in \eqref{barhi}, and there exist $0<\lambda_i,\gamma <2\alpha-\nu_1, \beta_i>0$, $i=1,\cdots,n-1$, such that
\begin{IEEEeqnarray}{rCl}
  \sup_{u}&\bigg[ \mathcal{G}_0+\gamma\mathcal{F}_0-\bigg(|\mathcal{G}_n |\chi^n+\sum_{i=1}^{n-1}|\mathcal{G}_i+\gamma\mathcal{F}_i|\chi^i\bigg)\nonumber\\
  &+\frac{\pa \Q_{n-1}}{\pa x_n}u\bigg]\geq 0, \label{recursivecbf2}
\end{IEEEeqnarray}
where $\chi=\omega_0+\sqrt{\omega_0^2+\frac{\omega_1^2}{2\nu_1\kappa}}$ and $\mathcal{F}_i$, $\mathcal{G}_i$ are defined in \eqref{dothpoly}. Then any Lipschitz continuous controller  
$$
u(x) \in K_{DOB}^{re}(x,\hat{d})\triangleq\left\{u\mid \psi_{0}^{re}+\psi_{1}^{re} u\geq 0\right\}
$$  
where
\begin{subequations}\label{recursivecbf}
\begin{align}
\psi_{0}^{re}&=\mathcal{P}(\bar x_n,\hat d)+\gamma\Q_{n-1}(\bar x_n,\hat d),\\
\psi_{1}^{re}&=\frac{\pa \Q_{n-1}}{\pa x_n} ,    
\end{align}
\end{subequations}
will guarantee $h(x_1(t))\geq 0$ for all $t\geq 0$.
\end{theorem}

\begin{proof}
We will show that $\bar h_i\geq 0$ indicates $\bar h_{i-1}\geq 0$ for any $t>0$ if $\bar h_{i-1}(\bar x_i(0),0)\geq 0$, $i=1,2,\cdots,n-1$.

\subsubsection*{Step 1}
Note that $\dot{h} +\lambda_1 h$ satisfies
\begin{IEEEeqnarray}{rCl}
\dot{h} +\lambda_1 h
    &=& \ph (f_1+\hat d) - \ph e_d +\lambda_1 h\nonumber\\
    &\geq& \ph (f_1+\hat d)-\frac{1}{2\beta_1}\phb^2+\lambda_1h-\frac{\beta_1}{2}e_d^2\nonumber\\
    &=& \bar h_1(\bar x_2,t),\label{q0}
\end{IEEEeqnarray}
from which it can be seen that $\bar h_1\geq 0$ indicates $\dot{h} +\lambda_1 h\geq 0$; thus, $h\geq 0$ for any $t>0$ as $h(x_1(0))\geq 0$.  


\subsubsection*{Step $i$ ($i=2,\cdots,n-1$)}
It can be seen that $\dot{\bar{h}}_{i-1}+\lambda_{i}\bar h_{i-1}$ satisfies
\begin{IEEEeqnarray}{rCl}
    &&\dot{\bar{h}}_{i-1}+\lambda_{i}\bar h_{i-1}\nonumber\\
    &=& \frac{\pa \Q_{i-1}}{\pa x_1}(f_1+d)+\sum_{j=2}^{i}\frac{\pa \Q_{i-1}}{\pa x_j}f_j
    +\frac{\pa \Q_{i-1}}{\pa \hat d}\dot{\hat{d}}
    -\beta_{i-1}\dot V_1\nonumber\\
    &&+\lambda_{i}\left(\Q_{i-1}-\frac{\beta_{i-1}}{2}e_d^2\right)
    \nonumber\\
    &\stackrel{\eqref{dotv}}{\geq}& \frac{\pa \Q_{i-1}}{\pa x_1}(f_1+\hat d)+\sum_{j=2}^{i}\frac{\pa \Q_{i-1}}{\pa x_j}f_j+\lambda_{i}\Q_{i-1}-\frac{\beta_{i-1}\omega_1^2}{2\nu_1}\nonumber\\
    && -\bigg(\alpha\frac{\pa \Q_{i-1}}{\pa \hat d}+\frac{\pa \Q_{i-1}}{\pa x_1}\bigg)e_d+\beta_{i-1}\bigg(\alpha-\frac{\nu_1}{2}-\frac{\lambda_{i}}{2}\bigg) e_d^2\nonumber\\
    &\geq&\frac{\pa \Q_{i-1}}{\pa x_1}(f_1+\hat d)+\sum_{j=2}^{i}\frac{\pa \Q_{i-1}}{\pa x_j}f_j+\lambda_{i}\Q_{i-1}-\frac{\beta_{i-1}\omega_1^2}{2\nu_1}\nonumber\\
    &&+ \bigg[ \frac{\left(\alpha\frac{\pa \Q_{i-1}}{\pa \hat d}\!+\!\frac{\pa \Q_{i-1}}{\pa x_1}\right)}{2\sqrt{\beta_{i-1}\!\left(\alpha\!-\!\frac{\nu_1}{2}\!-\!\frac{\lambda_{i}}{2}\right)}}\!-\!\sqrt{\beta_{i-1}\!\left(\alpha\!-\!\frac{\nu_1}{2}\!-\!\frac{\lambda_{i}}{2}\right)}e_d \bigg]^2\nonumber\\
    &&-\frac{\left(\alpha\frac{\pa \Q_{i-1}}{\pa \hat d}+\frac{\pa \Q_{i-1}}{\pa x_1} \right)^2}{\beta_{i-1}\left(4\alpha\!-\!2\nu_1\!-\!2\lambda_{i}\right)}\nonumber\\
    &\geq& \Q_i(\bar x_{i+1},\hat d)
    ,\label{mathcalqi}
\end{IEEEeqnarray}
from which it can be seen that $\bar h_i\geq 0$ indicates $\Q_i\geq 0$ and $\dot{\bar{h}}_{i-1}+\lambda_i \bar h_{i-1}\geq 0$; thus,  $\bar h_{i-1}\geq 0$ for any $t>0$ because $\bar h_{i-1}(x_i(0),0)\geq 0$.  

\subsubsection*{Step $n$}
Similar to the steps above, one can see that $\dot{\bar{h}}_{n-1}$ satisfies
\begin{IEEEeqnarray}{rCl}
    \dot{\bar{h}}_{n-1}
    &\geq& \mathcal{P}(\bar x_n,\hat d)+\frac{\pa \Q_{n-1}}{\pa x_n}u+\frac{\beta_{n-1}\gamma}{2}e_d^2,\label{diffdotbarhn1}
\end{IEEEeqnarray}
Selecting $u\in K_{DOB}^{re}$ yields
$
    \dot{\bar{h}}_{n-1}\geq -\gamma\bar h_{n-1}
$,
which implies $\bar h_{n-1}\geq 0$ for any $t>0$ as $\bar h_{n-1}(\bar x_n(0),0)\geq 0$. Therefore, one can conclude $h(x_1(t))\geq 0$ since $\bar h_i(x_{i+1}(0),0)\geq 0$, $i=0,1,\cdots,n-1$.
\end{proof}
The safe controller proposed in Theorem \ref{theorem5} is obtained by solving the following DOB-CBF-QP:%
\begin{align}
\min_{u} \quad & \|u-u_{nom}\|^2\label{cbfQP2}\\
\textrm{s.t.} \quad & \psi_0^{re}+\psi_1^{re} u\geq 0, \nonumber\\
&\mbox{DOB given in} \; \eqref{dobmismatched1d},\nonumber
\end{align}
where $\psi_0^{re},\psi_1^{re}$ are given in \eqref{recursivecbf} and $u_{nom}$ is any given nominal control law.


\subsubsection{Extended DOB-based Method}
\label{sec:quasilinear}
Now we will present an alternative approach to solve Problem \ref{prob2} based on the extended DOB. Compared with the DOB-CBF-QP controllers by Theorem \ref{theorem5},  controllers obtained from this second approach are often less conservative in simulations (see Section \ref{sec:simulation}).


Consider the system \eqref{system1dmismatched} and the safe set defined in \eqref{setcmismatched}.
Similar to \eqref{liftedh}, a set of functions $w_k(\bar x_{k+1},t), k=1, \cdots,n-1$, are defined as
\begin{equation}\label{liftedh1}
w_{k}(\bar x_{k+1},t)=\bigg(\frac{\diff}{\diff t}+\la_k\bigg)\circ w_{k-1}(\bar x_k,t),
\end{equation}
where $w_0(x_1,t)=h(x_1)$ and $\lambda_k>0$. Suppose that Assumption \ref{assumption2} holds with $r=n-1$, i.e., $\|d_E(t)\|\leq \xi_0$ and $|d^{(n)}|\leq \xi_1$, where $d_E(t)$ is defined in \eqref{deq}. We design the following extended DOB as in  \eqref{edob:both}:
\begin{IEEEeqnarray}{rCl}
\IEEEyesnumber\label{edob1:both} 
\IEEEyessubnumber
\hat d^{(i)} &=& p_{i}+l_{i}x_1,\label{edob1:sub1}\\
\IEEEyessubnumber \dot p_{i} &=& - l_{i}(f_1+\hat d^{(0)})+\hat d^{(i+1)}, \label{edob1:sub2}
\end{IEEEeqnarray}
where $\hat d^{(n)}(t)=0$ for any $t\geq 0$ and $\hat d^{(i)}_j$ denotes the estimate of $d^{(i)}_j$,  $i=0,1,\cdots,n-1$. Suppose that there exist functions $\mathcal{T}_0(\bar x_n)$, $\mathcal{T}_1(\bar x_n)$, $\mr_0(\bar x_n)$, $\mr_1(\bar x_n)$, $\mr_2(\bar x_n)$, such that
\begin{IEEEeqnarray}{rCl}
\IEEEyesnumber\label{quasiecbf:both} 
\IEEEyessubnumber
w_{n-1} \!&=&\! \mathcal{T}_0(\bar x_n)\!+\!\mathcal{T}_1(\bar x_n) d_E(t),\label{quasiecbf:sub1}\\
\IEEEyessubnumber
\dot{w}_{n-1}\!&=&\! \mr_0(\bar x_n)\!+\!\mr_1(\bar x_n) u\! +\!\mr_2(\bar x_n) d_E(t),\label{quasiecbf:sub2}
\end{IEEEeqnarray}
where $w_{n-1}$ is defined in \eqref{liftedh1}. Intuitively, condition \eqref{quasiecbf:both}  indicates that $w_{n-1}$ and $\dot w_{n-1}$ can be represented as the linear combination of $d$ and its derivatives with fixed $\bar x_n$. The following example about underactuated robotic systems further explains the idea of condition \eqref{quasiecbf:both}. 

\begin{example}\label{exmoti}
The underactuated robotic system given in \cite{huang2018disturbance} can be represented as follows after applying Olfati’s global transform of coordinates:
\begin{equation}\label{underactuatedrobot}
\begin{split}
    \dot x_1 &= m_{11}^{-1}(x_3) x_2,\\
    \dot x_2 &= f_1(x_1,x_3)+d_1,\\
    \dot x_3 &= x_4,\\
    \dot x_4 &= f_2(x_1,x_2,x_3,x_4)+b(x_1,x_2,x_3,x_4)u+d_2,
\end{split}
\end{equation}
where $x_1, x_2, x_3, x_4\in\R$ denote the state variables, $u\in\R$ is the control input, and $d_1,d_2\in\R$ are external disturbances. It is easy to verify that the disturbance relative degree of $d_1$ is lower than the input relative degree. Although system \eqref{underactuatedrobot} is slightly different from \eqref{system1dmismatched}, condition \eqref{quasiecbf:both} can still be satisfied: $w_2$ defined in \eqref{liftedh1} and its derivative $\dot{w}_2$ can be expressed as
\begin{IEEEeqnarray}{rCl}
\IEEEyesnumber \label{barh:both}
\IEEEyessubnumber \label{barh}
w_2&=&\ms_1 x_4+\ms_2
d_1+\ms_3,\\
\IEEEyessubnumber\label{dotbarh}
\dot{w}_2&=&\mathcal{V}_1+\mathcal{V}_2d_1+\ms_2\dot d_1+\ms_1 d_2+\ms_1bu,
\end{IEEEeqnarray}
where $\ms_1\!= \!\ph \frac{\pa m_{11}^{-1}}{\pa x_3}x_2$, $ \ms_2= \ph m_{11}^{-1}$, $\ms_3\! =\! \frac{\pa^2 h}{\pa x_1^2}(m_{11}^{-1}\!x_2)^2\! +\!\ph m_{11}^{-1}(f_1\!+\!\lambda_1x_2\!+\!\lambda_2x_2)\!+\!\lambda_1\lambda_2h$, $\mathcal{V}_1 = \left( \frac{\pa \ms_1}{\pa x_1}x_4+\frac{\pa \ms_2}{\pa x_1}+\frac{\pa \ms_3}{\pa x_1}\right)m_{11}^{-1}x_2+\ms_1f_2+\left(\frac{\pa \ms_1}{\pa x_3}x_4\!+\!\frac{\pa \ms_3}{\pa x_3}\right)x_4\!+\!\left(\frac{\pa \ms_2}{\pa x_2}x_4\!+\!\frac{\pa \ms_2}{\pa x_2}\!+\!\frac{\pa \ms_3}{\pa x_2}\right)f_1$, and $\mathcal{V}_2 = \frac{\pa \ms_2}{\pa x_2}x_4+\frac{\pa \ms_2}{\pa x_2}+\frac{\pa \ms_3}{\pa x_2}$.
Note that \eqref{barh:both} is equivalent to \eqref{edob1:both}.
Thus, it can be seen that for the system given in \eqref{underactuatedrobot}, any $h\in C^2$ satisfies condition \eqref{quasiecbf:both}.
\end{example}


Based on the extended DOB given in \eqref{edob1:both} and the condition shown in \eqref{quasiecbf:both}, the following result provides a QP-based safe controller for solving Problem \ref{prob2}.
 
\begin{theorem}\label{theorem4}
Consider the system \eqref{system1dmismatched}, the safe set $\tilde\C$ defined in \eqref{setcmismatched}, and the extended DOB given in \eqref{edob1:both} with $\hat d_E(0)=0$. Suppose that  Assumption \ref{assumption2} holds with $r=n-1$, $w_k(\bar x_{k+1}(0),0)>0$, $k=0,1,\cdots,n-1$, and condition \eqref{quasiecbf:both} holds. Furthermore, suppose that 
there exist $0<\gamma<2\kappa_E$, $\beta>\frac{\|E_d(0)\|^2}{2w_{n-1}(\bar x_n(0),0)}$, such that
\begin{IEEEeqnarray}{rCl}
   \sup_u &&\bigg[\mr_0+\gamma \mathcal{T}_0 -\|\mr_2+\gamma \mathcal{T}_1\|\chi_E-\frac{\beta\xi_1^2}{2\nu_2}+\mr_1 u\nonumber\\
&&-\frac{(\mr_2+\gamma\mathcal{T}_1)^2}{\beta(4\kappa_E-2\gamma)} \bigg]\geq 0,\label{quasiecbfcondition}
\end{IEEEeqnarray}
where $\kappa_E$ is defined in \eqref{edobdotv} and $\chi_E=\xi_0+\sqrt{\xi_0^2+\frac{\xi_1^2}{2\nu_2\kappa_E}}$. 
Then any Lipschitz continuous controller  
$$
u(x) \in K_{EDOB}(x,\hat{d}_j^{(i)})\triangleq\left\{u\mid \psi_{0}^E+\psi_{1}^E u\geq 0\right\}
$$  
will guarantee $h(x_1(t))\geq 0$ for all $t\geq 0$, where
\begin{subequations}\label{kcbfrnew}
\begin{align}
\psi_{0}^E\!&=\mr_0+\gamma \mathcal{T}_0 +(\mr_2+\gamma \mathcal{T}_1)\hat d_E\nonumber\\
&-\frac{(\mr_2+\gamma\mathcal{T}_1)^2}{\beta(4\kappa_E-2\gamma)} -\frac{\beta\xi_1^2}{2\nu_2},\\
\psi_{1}^E\!&=\! \mr_1.  
\end{align}
\end{subequations}

\end{theorem}

\begin{proof}
 If $\hat d_E(0) = 0$, then $\hat d_E(t)$ satisfies
$\|\hat d_E(t)\|=\|d_E(t)+\tilde e_d(t)\|\leq \chi_E$ from \eqref{bounded}.
Define a new CBF candidate $\bar h_E$ as:
\begin{equation}
    \bar h_E=  w_{n-1} -\beta V_2,
\end{equation}
where $V_2=\fot \|\tilde{d}_E\|^2$. Since 
\begin{IEEEeqnarray}{rCl}
   \dot{\bar{h}}_E &\stackrel{\eqref{edobdotv}}{\geq}& \mr_0+\mr_1 u +\mr_2 d_E+\beta\kappa_E\|\tilde e_d\|^2-\frac{\beta\xi_1^2}{2\nu_2}\nonumber\\
   &=& \mr_0\!+\!\gamma \mathcal{T}_0\!+\!\mr_1 u \!+\!(\mr_2\!+\!\gamma \mathcal{T}_1)\hat d_E\!-\!\gamma(\mathcal{T}_0\!+\!\mathcal{T}_1 d_E)\nonumber\\
   &&-(\mr_2+\gamma\mathcal{T}_1)\tilde e_d
   +\beta\kappa_E\|\tilde e_d\|^2-\frac{\beta\xi_1^2}{2\nu_2}
   \nonumber\\
    &\stackrel{\eqref{quasiecbf:both}}{\geq}& \mr_0+\gamma \mathcal{T}_0+\mr_1 u +(\mr_2+\gamma \mathcal{T}_1)\hat d_E-\gamma w_{n-1}\nonumber\\
   &&+\bigg\|\sqrt{\beta\left(\kappa_E\!-\!\frac{\gamma}{2}\right)} E_d-\frac{(\mr_2\!+\!\gamma\mathcal{T}_1)}{2\sqrt{\beta\left(\kappa_E\!-\!\frac{\gamma}{2}\right)}} \bigg\|^2\!\nonumber\\
   &&-\frac{(\mr_2+\gamma\mathcal{T}_1)^2}{\beta(4\kappa_E-2\gamma)}+\frac{\gamma\beta}{2}\|\tilde e_d\|^2-\frac{\beta\xi_1^2}{2\nu_2}\nonumber\\
   &\geq& \mr_0+\gamma \mathcal{T}_0+\mr_1 u +(\mr_2+\gamma \mathcal{T}_1)\hat d_E-\gamma w_{n-1}\nonumber\\
   &&-\frac{(\mr_2+\gamma\mathcal{T}_1)^2}{\beta(4\kappa_E-2\gamma)}+\frac{\gamma\beta}{2}\|\tilde e_d\|^2-\frac{\beta\xi_1^2}{2\nu_2},
\end{IEEEeqnarray}
selecting $u\in K_{EDOB}$  yields
\begin{equation}
    \dot{\bar{h}}_E\geq -\gamma w_{n-1}+\frac{\gamma\beta}{2}\|\tilde e_d\|^2=-\gamma \bar h_E.\label{generaldoth11}
\end{equation}
Note that the selection of $\beta$ ensures $\bar h_E(\bar x_n(0),0)\geq 0$. Thus, \eqref{generaldoth11} implies $\bar h_E\geq 0$ and $w_{n-1}\geq 0$ for any $t>0$. As $w_k(\bar x_{k+1}(0),0)\geq 0$, $k=0,1,\cdots,n-1$, it easy to see that $h(x_1(t))\geq 0$ for any $t>0$.
\end{proof}


The safe controller proposed in Theorem \ref{theorem4} is obtained by solving the following DOB-CBF-QP:%
\begin{align}
\min_{u} \quad & \|u-u_{nom}\|^2\label{cbfQP3}\\
\textrm{s.t.} \quad & \psi_0^{E}+\psi_1^{E} u\geq 0, \nonumber\\
&\mbox{extended DOB given in} \; \eqref{edob1:both},\nonumber
\end{align}
where $\psi_0^{E},\psi_1^{E}$ are given in \eqref{kcbfrnew} and $u_{nom}$ is any given nominal control law.

\section{Simulation Examples}
\label{sec:simulation}
In this section, two examples are presented to illustrate the effectiveness of the proposed methods. The robust CBF method proposed in  \cite{nguyen2021robust} is used for comparison.

\begin{figure}[!htbp]
\centering
\begin{subfigure}{0.43\textwidth}
  \centering
  \includegraphics[width=1\textwidth]{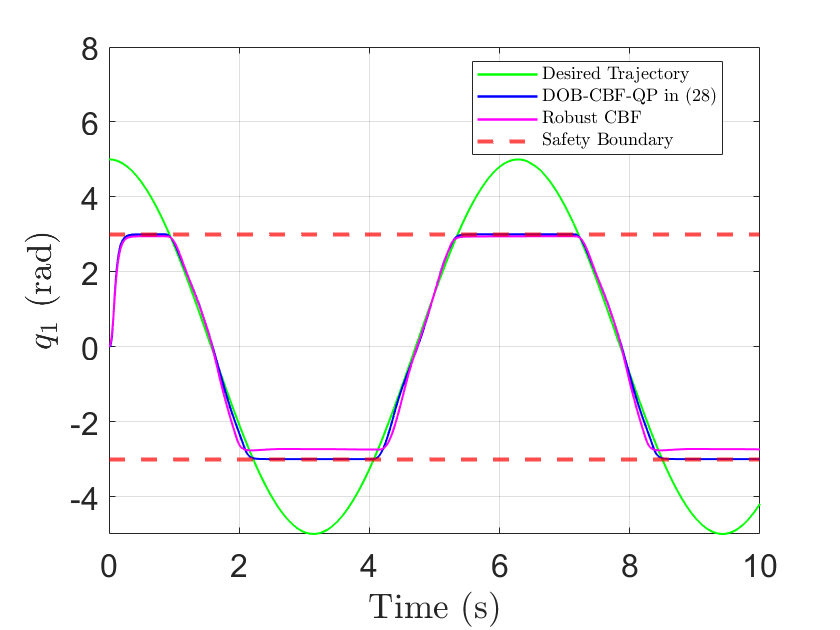}
  \caption{Evolution of the joint angle $q_1$}
\end{subfigure}
\begin{subfigure}{0.43\textwidth}
  \centering
  \includegraphics[width=1\textwidth]{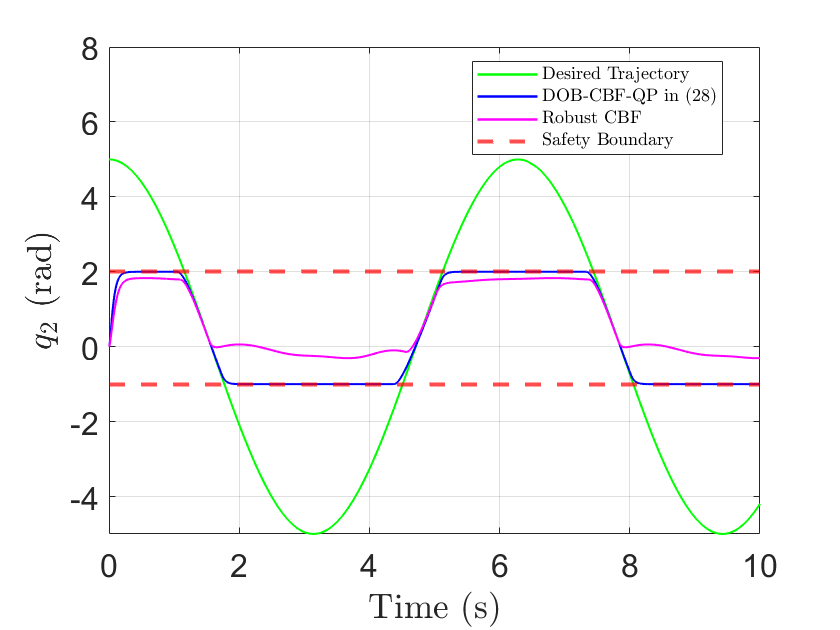}
  \caption{Evolution of the joint angle $q_2$}
\end{subfigure}
\caption{Simulation results of Example \ref{example1}. The trajectories of the closed-loop system with the proposed DOB-CBF-QP-based controller by solving \eqref{cbfQP1r} are always safe (i.e., $q_1,q_2$ are always inside the respective safe regions bounded by the dashed red lines) and are less conservative than the robust CBF approach in  \cite{nguyen2021robust} because the trajectories are able to track the desired trajectories (i.e., the green lines) much better inside the safe region.}
\label{fig:robot}
\end{figure}

\begin{example}\label{example1}
Consider a 2-DOF planar robot whose dynamics are described by 
\begin{equation}
    M(q)\ddot q+C(q,\dot q)\dot q+G(q)=\tau+\tau_d,
\end{equation}
where $q=[q_1 \ q_2]^\top\in\R^2$ denote the joint angles, $\tau\in\R^2$ is the control input, and $\tau_d\in\R^2$ represents the external disturbance satisfying $\|\tau_d\|\leq 30$ and $\|\dot\tau_d\|\leq 50$. For the robust CBF approach, the bound of $\tau_d$ is also set as $30$. 
The physical parameters used are chosen as those in \cite{sun2011neural}. The following four CBFs are employed to represent the constraints on joint angles: $h_1=q_1+3$, $h_2=3-q_1$, $h_3=q_2+1$, and $h_4=2-q_2$. Clearly, the input relative degree of the system is equal to the disturbance relative degree. It can be verified that the conditions of Theorem \ref{theorem2} hold, so that a DOB-CBF-QP-based controller can be obtained by solving \eqref{cbfQP1} to ensure the safety of the closed-loop system. The simulation results are presented in Fig. \ref{fig:robot}. It can be seen that the trajectories of the closed-loop system with the proposed DOB-CBF-QP-based controller by solving \eqref{cbfQP1} are always safe as $q_1,q_2$ stay inside the respective safe regions bounded by the dashed red lines; furthermore, the  trajectories of the closed-loop system are less conservative than the robust CBF approach because the trajectories are able to track the desired trajectories (i.e., the green lines) much better inside the safe region.
\end{example}

\begin{figure}[htbp]
\centering
\begin{subfigure}{0.43\textwidth}
  \centering
  \includegraphics[width=1\textwidth]{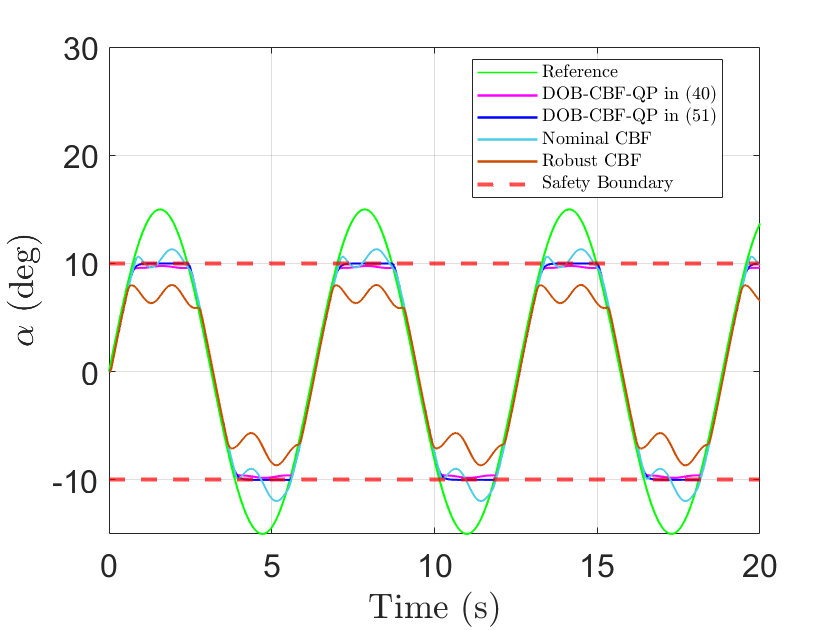}
  \caption{Evolution of the angle of attack $\alpha$ by using two CBFs $h_1=10-\alpha$ and $h_2=\alpha+10$}
\end{subfigure}
\begin{subfigure}{0.43\textwidth}
  \centering
  \includegraphics[width=1\textwidth]{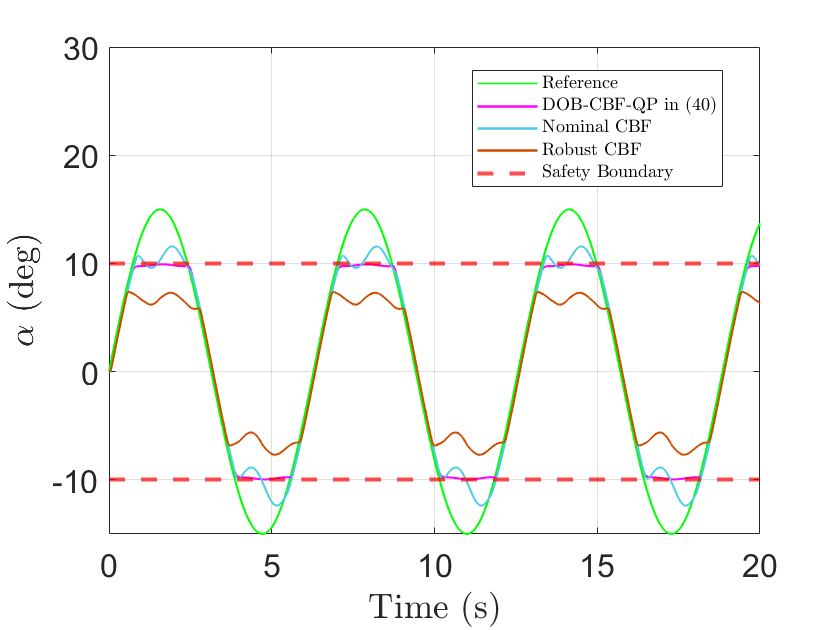}
  \caption{Evolution of the angle of attack $\alpha$ by using a single CBF $h=100-\alpha^2$}
\end{subfigure}
\caption{Simulation results of Example \ref{example2}. It can be seen that the proposed DOB-CBF-QP-based controllers in \eqref{cbfQP2} and \eqref{cbfQP3}
are both less conservative than the robust CBF approach in \cite{nguyen2021robust} because the trajectories are able to track the desired trajectories (i.e., the green lines) much better inside the safe region. Moreover, from (a) one can see that the DOB-CBF-QP from \eqref{cbfQP3} has a better tracking performance than that from \eqref{cbfQP2}.}
\label{fig:missile}
\end{figure}

\begin{example}\label{example2}
Consider the longitudinal dynamics of a missile given in \cite{yang2013static}:
\begin{IEEEeqnarray}{rCl}
\IEEEyesnumber \label{missile:both}
\IEEEyessubnumber
\dot\alpha &=& f_1(\alpha) +q+b_1(\alpha)\delta+d_1,\\
\IEEEyessubnumber \dot q &=& f_2(\alpha) + b_2\delta +d_2,\\
\IEEEyessubnumber \dot\delta &=& (1/t_1)(-\delta+u),
\end{IEEEeqnarray}
where $\alpha$ is the angle of attack, $q$ is the pitch rate, $\delta$ is the tail fin deflection, $u$ is the control input,  $d_1$, $d_2$ are mismatched disturbances satisfying $\|d_1\|\leq 30$, $\|d_2\|\leq 30$, $\|\dot d_1\|\leq 50$, $\|\dot d_2\|\leq 50$, and $f_1(\alpha)$, $b_1(\alpha)$, $f_2(\alpha)$ are nonlinear known functions whose explicit forms are given in \cite{yang2013static}. The same bounds above are used for the robust CBF approach. It  is obvious that the disturbance relative degree of $d_1$ is lower than the input relative degree. Although system \eqref{missile:both} is slightly different from \eqref{system1dmismatched}, the control approaches presented in Section \ref{sec:mismatcheddobre} can be easily generalized to deal with such a system.

To avoid the stall,
the proposed DOB-CBF-QP-based controllers are employed to restrict the value of $\alpha$ in the presence of disturbances. We consider two scenarios: 1) Two CBFs are selected as $h_1=10-\alpha$ and $h_2=\alpha+10$; in this case, the DOB-CBF-QP-based controllers both in \eqref{cbfQP2} and \eqref{cbfQP3} are applicable. 2) A single quadratic CBF $h=100-\alpha^2$ is employed; in this case, only the DOB-CBF-QP-based controller in \eqref{cbfQP2} is applicable. From the simulation results shown in Fig. \ref{fig:missile}, it can be seen that the proposed safe controllers can both ensure safety (i.e., the trajectory of $\alpha$ is always inside the safe regions bounded by the dashed red lines), and the trajectories of the closed-loop system are less conservative than the robust CBF approach because of the better tracking performance inside the safe region. Meanwhile, it can be observed that the DOB-CBF-QP-based controller obtained from \eqref{cbfQP3} has a better tracking performance than that from \eqref{cbfQP2}.
\end{example}

\section{Conclusion}
\label{sec:conclusion}
In this paper, a new DOB-CBF-QP-based safe control design approach
was proposed for systems with external disturbances, with the goal of achieving a better tradeoff between safety and performance. 
The simulation results demonstrate the superiority of the proposed control scheme over existing robust CBF techniques. Future studies include applying the proposed control technique to systems with measurement uncertainties and relaxing the assumptions of the theoretical results. 




\bibliographystyle{IEEEtran}
\bibliography{ACC23-DOB} 

\end{document}